*Astro2010 State of the Profession Position Paper*
# Research Science and Education: The NSF's Astronomy and Astrophysics Postdoctoral Fellowship
**March 2009**

Dara Norman (NOAO), Marcel Agueros (Columbia Univ), Timothy M. Brown (LCOGT), Matthew Browning (UC, Berkeley), Sukanya Chakrabarti (Harvard CfA), Bethany Cobb (UC, Berkeley), Kim Coble (Chicago State Univ.), Christopher Conselice (Univ. of Nottingham), Kelle Cruz (Caltech), Laura Danly (Griffith Obs.), Peter M. Frinchaboy III (Univ of Wisconsin, Madison), Eric Gawiser (Rutgers Univ.), Joseph Gelfand (NYU), Anthony Gonzalez (Univ. of Florida), Jennifer L. Hoffman (Univ. of Denver), Dragan Huterer (Univ. of Michigan), John Johnson (Univ. of Hawaii), Roberta M. Johnson (NCAR). Sheila Kannappan (UNC), Rachel Kuzio de Naray (UC, Irvine), David Lai (UC, Santa Cruz), Douglas C. Leonard (San Diego State Univ.) Makenzie Lystrup (Univ. of Colorado, Boulder), Sera Markoff (Univ. of Amsterdam), Karín Menéndez-Delmestre (OCIW), Stephan Muchovej (Caltech), M. Virginia McSwain (Lehigh University), Katherine Rhode (Indiana Univ.), Tammy Smecker-Hane (UC, Irvine) Malcolm Smith (CTIO), Jennifer Sokoloski (Columbia Univ.), Kim-Vy Tran (Texas A&M)

Primary Contact
Dara Norman
NOAO
950 N. Cherry Ave
Tucson, AZ 85719
dnorman@noao.edu,
520-318-8361

*The Astronomy and Astrophysics Postdoctoral Fellowship*

**Abstract**
The NSF's Astronomy and Astrophysics Postdoctoral Fellowship (AAPF) is exceptional among the available postdoctoral awards in Astronomy and Astrophysics. The fellowship is one of the few that allows postdoctoral researchers to pursue an original research program, of their own design, at the U.S. institution of their choice. However, what makes this fellowship truly unique is the ability of Fellows to lead an equally challenging, original educational program simultaneously. The legacy of this singular fellowship has been to encourage and advance leaders in the field who are equally as passionate about their own research as they are about sharing that research and their passion for astronomy with students and the public.

In this positional paper we address the importance of fellowships like the AAPF to the astronomical profession by identifying the science and educational contributions that Fellows have made to the community. Further, we recommend that fellowships that encourage leading postdoctoral researchers to also become leaders in Astronomy education be continued and expanded.

## 1) What is the AAPF?

The basic description of the Fellowship is best summarized in the NSF's program synopsis. We reproduce it here: The "Astronomy and Astrophysics Postdoctoral Fellowships provide an opportunity for highly qualified, recent doctoral scientists to carry out an integrated program of independent research and education. Fellows may engage in observational, instrumental, theoretical, laboratory or archival data research in any area of astronomy or astrophysics, in combination with a coherent educational plan for the duration of the fellowship. The program supports researchers for a period of up to three years with fellowships that may be taken to eligible host institution(s) of their choice. The program is intended to recognize early-career investigators of significant potential and to provide them with experience in research and education that will establish them in positions of distinction and leadership in the community."[1]

While this synopsis is descriptive, below we point out the combination of elements of this fellowship that make it truly unique among all others offered in Astronomy or Astrophysics.

### a. Postdoctoral fellowship to pursue independent research

The fellowship provides a realistic simulation of the mix of responsibilities that are given to university faculty and has allowed the AAPFs to demonstrate that postdoctoral researchers are capable of simultaneous excellence in research and education activities. The AAPF program produces alumni who have successfully navigated the NSF grant process from start to finish. It offers an important alternative to other prize fellowship programs where letters of recommendation are given high priority; the AAPF grant rewards those who are scientifically mature enough at this stage of career to propose and conduct frontier independent research. The vagaries and potential abuses represented by confidential letters of recommendation[2] make it critical for the profession to have prestigious career paths available for people who have greater talent than their letters might imply.

---

[1] Full program details can be found at
www.nsf.gov/funding/pgm_summ.jsp?pims_id=5291

### b. Research can be done in any astronomical/astrophysical subfield

Recipients of this fellowship have pursued research in a wide variety of astronomical subfields. Work has ranged from instrumentation to theoretical to observational, across wavelengths from X-ray to radio and over source scales from planetary to large-scale structure. In the section below, there are summaries of a number of outstanding research projects in which Fellows have leading roles. As a result of the breadth of these fields, AAPF Fellows have gone on to careers at a wide range of institutions, including research universities, teaching universities, and national observatories and laboratories (see the table below).

### c. Fellowship can be taken to any U.S. institution

Fellows are required to justify how their proposed host institution can support both their research and education/outreach program. Thus fellowships have been hosted by a wide variety of universities and national observatories (see table below) that not only commit to the scientific program that the Fellow has proposed, but also to the educational program proposed. The flexibility of the host institution selection allows Fellows to address both academic and personal objectives.

With regard to academic objectives, Fellows have been able to leverage host connections in order to begin domestic and international collaborations both in research and in education. In many cases, host institutions have become more engaged in education and outreach to the larger community through the connections that Fellows have cultivated.

In addition, the flexibility of this fellowship is of particular importance given that "two-body problems" and family relocation issues that affect many astronomers. In many cases the NSF AAPF has allowed excellent researchers – who, because of limited career opportunities in a particular geographic location, might have had to leave astronomy at the postdoctoral stage, or compromise their careers in some other way – to instead take the fellowship to the institution of their choice, continue to successfully contribute to the field, and use the AAPF as a stepping stone to a permanent position.

### d. An educational program is a __major__ component of the proposal.

The AAPF is the __only__ prize fellowship that requires recipients to commit a substantial percentage of their time to a deliberate, planned education or public outreach project of their own design. This component has been interpreted broadly and as a result, Fellows have engaged in a full spectrum of activities from direct formal teaching initiatives and curriculum development to unique, new approaches to community outreach. These education and outreach programs have included work with minority students and teachers in underserved areas, the incorporation of new teaching/learning techniques into astronomy/physics classes and international teacher/student collaborations.
The AAPF is committed to encouraging, supporting and promoting excellent researchers who have interests and experience in education and public outreach.

---

[2] Trix, F. & Psenka, C., "Exploring the color of glass: Letters of recommendation for female and male medical faculty", Discourse & Society 14(2003): 191-220.

## 2) Astronomy Researchers and Educators

The 2001 astronomy decadal review commissioned by the NRC, entitled "Astronomy and Astrophysics in the New Millennium", presented recommended priorities for the most important new initiatives to be accomplished during the 2000-2010 decade. While much of the document focuses on key science goals and large projects, there is a substantial section that addresses the role of astronomy in education. The final section of the executive summary gives recommendations for the participation of astronomers in education.[3] The committee recommended that 1) Astronomers be involved in science education at the K-12 level with both students and teachers, as well as with the general public, 2) the value of this work be recognized by the astronomical community, 3) federal agencies provide incentives for astronomers to be more engaged in education, and 4) the NSF invest resources to improve public recognition of the achievements of astronomy.

The AAPF satisfies important aspects of all of these recommendations. Through the fellowship, the NSF has provided funds for a program that allows Fellows to carry out projects that engage them as professional researchers in a spectrum of education and public outreach activities that increase public recognition of astronomical achievement. It is through this combination of excellence as researchers and educators that the value of this work has been recognized in the community.

## 3) Recommendations
- The AAPF should continue to be funded by the NSF at a level that makes it even more desirable than other postdoctoral fellowships since this program uniquely prepares researchers for careers as leaders in the field, at a variety of institutions.
- The AAPF should be expanded to allow more postdoctoral researchers the opportunity to advance their careers as researchers, teachers and mentors.
- Universities should create their own pre-faculty versions of this post-doctoral experience. This idea is not without precedent: A few examples of these fellowships already exist, e.g. the Presidential Fellows program in the UC system and the Carolina Postdoc program for faculty diversity at UNC.

## 4) Meet the AAPF Fellows

Over the nearly 8 year span of this fellowship there have been 65 Astronomy and Astrophysics Postdoctoral Fellows. **To date, 80% of alumni (2001-2005) hold professor/ lecturer/scientist positions or higher.** The required brevity of this document precludes including the details of all their research and educational works. Below we highlight results from a wide selection of programs undertaken by Fellows that span the breadth of science, education and outreach activities. Additional information about all the Fellows and the AAPF can be found on the website, maintained entirely by the Fellows, at http://aapf-fellows.org

---

[3] See NRC's "Astronomy and Astrophysics in the New Millennium", pages 47-49.

### a. Selected Examples of AAPF Contributions to Astronomical and Astrophysical Research

**Simulations Explore Solar Magnetism:** Fellow Matthew Browning and collaborators published the first global numerical simulations of the solar dynamo that include penetration into the tachocline of shear. Their computational models solved the full three- dimensional equations of fluid motion and magnetism using massively parallel supercomputers. In their calculations, the tachocline is found to play a major role in generating the magnetism, yielding mean (large-scale) fields that are much stronger and more organized than in prior simulations that modeled only the convective envelope. (Browning, M.K., Miesch, M.S., Brun, A.S., Toomre, J., 2006, ApJ 648, 157)

**Finding Distant Earths Faster:** Observing a planet called WASP- 10b, which orbits a star approximately 300 light years distant from Earth, a team of astronomers led by AAPF Fellow John Johnson, measured precisely how much the star dimmed as the planet passed in front of it. The team reported that they were able to detect a dimming of less than 0.05% during a single transit of 2.2 hours. In so doing, the team demonstrated the highest ground-based photometric precision ever attained for a single planet transit. "This instrument provides us with the first realistic chance to detect a transiting Earth-like planet," said Johnson. "In the past, you'd have to spend maybe 10 nights on the telescope to get this precision, but now we are doing it in a single night." (Johnson et al, 2009 ApJ 692 100)

**How Be Stars Get Their Spin:** Be stars are a class of rapidly rotating stars surrounded by circumstellar disks of hot material. The stars may have been born as rapid rotators or they may have been spun up later either by mass transfer from a binary companion or during the evolution of normal stars. To test these three scenarios, AAPF Fellow M. Virginia McSwain at Yale University and her collaborator Doug Gies at Georgia State University conducted a survey of Be stars in 55 stellar clusters in the southern sky. They found a surprising number of older Be stars in these clusters. Consequently, a spin-up phase during normal stellar evolution cannot produce the observed age distribution of Be stars, but up to 75% of the Be stars detected may have been spun up by binary mass transfer. Most of the remaining Be stars were likely rapid rotators at birth. (McSwain, M. V., & Gies, D. R. 2005, ApJS, 161, 118)

**An X-Ray Jet from a Nearby Star:** X-Ray Observations of a Type Ia Supernova Progenitor. Early in 2006, an eruption of the recurrent nova RS Ophiuchi provided the first opportunity to perform comprehensive X-ray observations of a nova embedded within a wind nebula and to diagnose conditions within the ejecta. AAPF Fellow Jennifer Sokoloski and collaborators showed that the hard X-ray emission from RS Ophiuchi early in the eruption emanated from behind a blast wave that expanded freely for less than 2 days and then decelerated due to interaction with the nebula. The X-rays faded rapidly, suggesting that the blast wave deviates from the standard spherical shell structure. The early onset of deceleration indicates that the ejected shell had a low mass, the white dwarf has a high mass, and that RS Ophiuchi is a progenitor of a type Ia supernova (Sokoloski et al. 2006, Nature, 442, 276).

**First Black Hole Discovered in a Globular Star Cluster:** Fellow Katherine Rhode and colleagues have discovered the first clear case of black hole in a globular cluster. As part of a survey of globular cluster systems of galaxies out to 20 Mpc, Rhode identified ~1,500 globular clusters in the luminous Virgo elliptical NGC4472. When Rhode and Tom Maccarone (Univ. of Southampton), Arunav Kundu and Stephen Zepf (Michigan State Univ.) observed NGC4472 with the X-ray Multiple Mirror-Newton telescope, they found a luminous, highly variable X-ray source within one of NGC4472's globular clusters. Archival Chandra and ROSAT observations confirmed the position and the persistence of this luminous X-ray source. This discovery yields new questions, such as why some globular clusters can harbor black holes while others don't, and whether there are parallels between supermassive black holes in galaxies and black holes in lower-mass systems like globular clusters. (Maccarone et al. 2007, Nature, 445, 183)

**MUSYC of the Spheres:** MUSYC (MUltiwavelength Survey by Yale-Chile, was created in 2003 as a merger of Eric Gawiser's funded fellowship proposal for a square-degree optical imaging plus spectroscopic survey with efforts at complementary wavelengths by faculty at Yale and U. de Chile. The 1.2-square-degree optical catalog contains 277,341 sources with a 50% completeness limit of R~26.5. The collaboration includes 30 investigators from the US, Chile, and Europe plus six Ph. D. students. All data from this survey will be made public, with reduced images and catalogs already available from the broad-band and narrow-band optical and near-IR imaging. The Survey has generated over 30 refereed publications including several major breakthroughs in our understanding of galaxy formation. (Gawiser et al. 2006 ApJS 162, 1)

**Galaxies As Magnifying Glasses:** The gravity of a galaxy can act as a giant distorting lens, producing optical illusions such as multiple images of background objects. Galaxies can produce these distorted images because their gravity bends and magnifies light rays coming from more distant objects. Einstein's theory of relativity predicts that these gravitational lensing phenomena should always produce an odd number of images, including a relatively faint image that appears near the center of the lens. Until now, however, none of these central images were ever seen. Fellow Josh Winn and his colleagues have recently reported the first known central image, which allows the team of astronomers to study the lensing galaxy and its central black hole. (Winn, J. Rusin, D., and Kochanek, C. 2004 Nature 427,613).

**Probing the Extragalactic Past:** AAPF Fellow Christopher Conselice has developed a new methodology to help determine how galaxies form and evolve. Distant galaxies have small sizes and low masses in comparison to modern galaxies. No one quite knows how these galaxies from the past evolved to become like those we see today, but many scientists suspect that galaxy mergers were an important part of the process. By developing new analysis tools that probe the galaxy structures observed in deep HST images, Conselice has shown that the most massive distant galaxies are undergoing major

*The Astronomy and Astrophysics Postdoctoral Fellowship*mergers. By considering the evolution of these mergers over time, Conselice has demonstrated that the most massive systems in the modern universe can indeed form from galaxy mergers—a result which beautifully confirms earlier suspicions. (Conselice, C.J., Bershady, M., Dickinson, M., Papovich, C. 2003, AJ, 126, 1183)

**Unexpected Correlations in the Cosmic Microwave Background:** A group of theoretical physicists now casts doubt on the standard interpretation of the WMAP data on the largest angular scales. Fellow Dragan Huterer and his colleagues have found that correlations in the temperature fluctuations vanish on scales above 60 degrees, and that the large-scale fluctuations themselves are aligned with the geometry and direction of motion of the solar system. The researchers estimate that the chances of these alignments being accidental are much less than part in a thousand. The origin of the correlations is unknown at this time and many possibilities, ranging from systematic errors to exotic particle physics explanations, are being investigated. (C.J. Copi, D. Huterer, D.J. Schwarz and G.D. Starkman, 2007 Phys. Rev. D 75, 023507)

### b. Selected Examples of AAPF Contributions to Astronomy Education

**Community College Students Reach for the Stars:** Hartnell College and the Center for Adaptive Optics (CfAO) at University of California-Santa Cruz (UCSC) offer a unique summer course, now entering its fourth year, designed to inspire Hartnell students to reach for the stars. Developed in 2004 by AAPF Fellow Anne Metevier, the intensive six-day course is taught by Hartnell instructors and astronomy graduate students from UCSC. It not only introduces students to basic astronomy, it gives them hands-on training in scientific research. Of the 47 students who have enrolled in the short course, 16 (34%) have transferred to four-year colleges to pursue science and engineering (S&E) degrees. Although the short course is open to all students, one of its goals is to bring more minorities into the sciences. According to a recent NSF survey, more than half (51%) of the Latino students who earn bachelors degrees in S&E fields attend community college first. Other races/ethnicities are not far behind: 45% of Native Americans, 44% of African Americans, 43% of Whites, and 40% of Asian Americans go to community college before obtaining S&E bachelors degrees. In conjunction with the record levels of minorities who are attending community colleges, these statistics highlight the important role of community colleges in the pipeline toward S&E careers. Consequently, outreach efforts directed toward community college students can be especially effective at encouraging minorities to pursue S&E professions.

**Collaborative Laboratory Course Enhances Student Learning:** Twenty percent of all undergraduates at the University of California, Los Angeles, are experiencing a new suite of laboratory modules that deepen their conceptual understanding of the physical principles of astronomy relative to their peers in traditional lab courses. The new laboratory course, developed by Fellow Nathan McCrady, is based upon research findings in education and cognitive science that show that learning is fundamentally influenced by context and has a strong social component. Consequently, the labs purposefully stimulate teamwork and discussion among student teams with inquiry-based, hands-on activities that apply a core set of astrophysical concepts repeatedly

- 6 -

*The Astronomy and Astrophysics Postdoctoral Fellowship*throughout the quarter in varying contexts and at increasing levels of sophistication. The new labs model "doing science:" what constitutes evidence, how experiment can be used to answer questions, how to build an argument from evidence, and how to work collaboratively. Thus, these labs develop an appreciation of the scientific process in an important constituency for public-funded science: future taxpayers.

### c. **Selected Examples of AAPF Contributions to Public Outreach**

**Source for the Latest News and Views on the Cosmos:** One of the strengths of astronomy is the wide variety of current research. The diversity of ongoing research is lost in most media coverage on Astronomy, which typically focuses on individual major discoveries. As a result, the breadth of current astronomy research, and the importance of fields not often covered in the new media, is often under-appreciated by the general public. To remedy this situation, Fellow Joseph Gelfand hosts a weekly, one-hour astronomy radio show where he describes the latest scientific results, ranging in subject matter from studies of Solar System objects to the latest cosmological surveys and simulations culled from press releases, scientific journals (for example Science and Nature), and the pre-print archive. The show focuses on the broader context of the science and explains the relevance to outstanding questions in each field. To supplement these topics, interviews with professional astronomers are regularly broadcast focusing on both the questions they are hoping to answer and the methods used to do so. On several occasions a series of shows has been devoted to a particular topic (e.g., galaxy formation) to provide a more complete description than a collection of press releases would allow. In addition to being broadcast live in the mid-Hudson region of NY state, recorded versions of each show and interview are posted on a website and several free podcast servers (e.g. iTunes). More than 5000 copies of radio shows and interviews have been downloaded in the 18 months this program has been on-air, demonstrating the intense interest for such material and the need to support such activities in the future.

**Educational Exchanges Across the Equator:** What is color? What is light? How can we use a spectrometer to help students understand the answers to these questions? Even half a world apart and between people of different languages and cultures, how to teach these ideas to students is a lively subject for discussion. Fellow Dara Norman and collaborators from NOAO North and South have sponsored teacher professional development videoconference workshops, dubbed ASTRO-Chile, linking middle school teachers in Tucson, AZ, and La Serena, Chile. The teachers exchange methods and ideas about how to explain and demonstrate physical concepts, important to the study of astronomy, to students of various ages. The workshops are conducted in Spanish with four bilingual science teachers from the Tucson area discussing pedagogical approaches with their teaching counterparts in Chile. Demonstrations and project presentations, from both sites, are a core part of each workshop.

### 5) **Other Fellow initiatives**
In addition to the AAPF website, the Fellows also independently host a symposium each year before the start of the winter AAS meeting to share their work with one another as well as hosting mentors, invited guests and the general astronomy community. These

- 7 -

*The Astronomy and Astrophysics Postdoctoral Fellowship*

symposia are enthusiastically attended and often lead to collaborations among the Fellows.

| Table 1: NSF AAPF Alumni and Current | | | |
|---|---|---|---|
| Year | Name | Fellowship institution | Current (or last known) position |
| 2001 | Kim Coble | Univ. of Chicago and Adler Planetarium | Asst. Prof., Dept. of Chemistry and Physics, Chicago State Univ. |
| 2001 | Kristy Dyer | National Radio Astronomy Obs., Socorro | NRC Fellow, Naval Research Lab |
| 2001 | Brenda Frye | Princeton Univ. | Lecturer, Dublin City Univ. (Ireland) |
| 2001 | Eric Hooper | Univ. of Texas, Austin | Asst. Prof., U. Wisconsin |
| 2001 | Denise Hurley-Keller | Case Western Reserve Univ. | Case Western Reserve Univ. |
| 2001 | Kelsey Johnson | Univ. of Wisconsin and NRAO, Socorro | Asst. Prof., Univ. of Virginia |
| 2001 | Brian Keating | Caltech | Asst. Prof., Center for Astrophysics & Space Sciences, Univ. of California, San Diego |
| 2001 | Dara Norman | Cerro Tololo InterAmerican Obs. | Asst. Scientist, NOAO |
| 2001 | Don Smith | Univ. of Michigan | Asst. Prof., Guilford College, Greensboro, NC |
| 2001 | Josh Winn | Harvard-Smithsonian CfA | Asst. Prof., MIT |
| 2002 | Philip Arras | Kavli Institute for Theoretical Physics (UCSB) | Asst. Prof., U. Virginia |
| 2002 | Christopher J. Conselice | Caltech | Assoc. Prof. & Reader Univ. of Nottingham |
| 2002 | Eric Gawiser | Yale Univ. | Asst. Prof., Rutgers Univ. |
| 2002 | Anthony Gonzalez | Univ. of Florida | Asst. Prof., Univ. of Florida |
| 2002 | Sera Markoff | MIT | Asst. Prof., Univ. of Amsterdam |
| 2002 | Jon M. Miller | Harvard-Smithsonian CfA | Asst. Prof., Univ. of Michigan |
| 2002 | Andrew Sheinis | Univ. of California at Santa Cruz | Asst. Prof., Univ. of Wisconsin |
| 2002 | Aparna Venkatesan | Univ. of Colorado, Boulder | Asst. Prof., U. San Francisco |

*The Astronomy and Astrophysics Postdoctoral Fellowship*

| Year | Name | Host Institution | Current Position |
|---|---|---|---|
| 2003 | Joseph Barranco | Kavli Institute for Theoretical Physics (UCSB) & Harvard-Smithsonian CfA | Asst. Prof., Dept. of Physics & Astronomy, San Francisco State Univ. |
| 2003 | John Feldmeier | Case Western Reserve Univ. and NOAO | Asst. Prof., Youngstown State |
| 2003 | Rose Finn | Univ. of Massachusetts | Asst. Prof., Siena College |
| 2003 | Jennifer L. Hoffman | Univ. of California at Berkeley | Asst. Prof. of Physics & Astronomy, Univ. of Denver |
| 2003 | Anne Metevier | Univ. of California, Santa Cruz | Lecturer, Sonoma State Univ. & Consultant, Center for Adaptive Optics Prof. Development Program |
| 2003 | Katherine Rhode | Wesleyan Univ. and Yale Univ. | Asst. Prof. of Astronomy, Indiana Univ. |
| 2003 | Jessica Rosenberg | Univ. of Colorado and Harvard-Smithsonian CfA | Asst. Prof., George Mason Univ. |
| 2003 | Jennifer Sokoloski | Harvard-Smithsonian CfA and Columbia Univ. | Assoc. Research Scientist, Columbia Astrophysics Laboratory |
| 2004 | Héctor G. Arce | American Museum of Natural History | Asst. Prof., Yale Univ. |
| 2004 | Kelle Cruz | American Museum of Natural History | Spitzer Fellow, California Institute of Technology |
| 2004 | Joshua Faber | Univ. of Illinois at Urbana-Champaign | Asst. Prof. of Mathematics, Rochester Institute of Technology |
| 2004 | Dragan Huterer | Univ. of Chicago | Asst. Prof. of Physics, Univ. of Michigan |
| 2004 | Sheila Kannappan | Univ. of Texas at Austin | Asst. Prof., Univ. of North Carolina |
| 2004 | Douglas C. Leonard | California Institute of Technology | Asst. Prof., Department of Astronomy, San Diego State Univ. |
| 2004 | M. Virginia McSwain | Yale Univ. | Asst. Prof., Lehigh Univ. |
| 2004 | Travis Metcalfe | High Altitude Obs., NCAR | Scientist, NCAR |
| 2004 | Henry Roe | California Institute of Technology | Asst. Astronomer, Lowell Obs. |
| 2005 | Jeffrey Bary | Univ. of Virginia | Asst. Prof., Colgate Univ. |
| 2005 | Matthew Browning | Univ. of California at Berkeley | Canadian Inst. For Theoretical Astrophysics |
| 2005 | Sukanya Chakrabarti | Harvard-Smithsonian CfA | ITC Fellow, Harvard-Smithsonian CfA |
| 2005 | Scott Dahm | Caltech | Keck Obs. |

- 9 -

*The Astronomy and Astrophysics Postdoctoral Fellowship*

| Year | Name | Institution | Current Position |
|------|------|-------------|------------------|
| 2005 | Daniel Kocevski | Univ. of California at Berkeley | |
| 2005 | Nathan McCrady | Univ. of California, Los Angeles | |
| 2005 | Erik Rosolowsky | Harvard-Smithsonian CfA | Asst. Prof., Univ. of British Columbia, Okanagan |
| 2005 | Kim-Vy Tran | Harvard-Smithsonian CfA | Asst. Prof., Texas A&M |
| 2006 | Marcel Agueros | Columbia | |
| 2006 | Peter Frinchaboy | Univ. of Wisconsin-Madison | |
| 2006 | Christopher Groppi | Univ. of Arizona, Steward Obs. | |
| 2006 | DeWayne Halfen | Univ. of Arizona, Steward Obs. | |
| 2006 | Thomas Renbarger | Univ. of California, San Diego | |
| 2006 | Tamara Rogers | UCAR | Asst. Prof., Univ. of Arizona |
| 2006 | David Rothstein | Cornell Univ. | |
| 2006 | Kurtis Williams | Univ. of Texas, Austin | |
| 2006 | Andrew Zentner | Kavli Institute for Cosmological Physics, Univ. of Chicago | Asst. Prof., Univ. of Pittsburgh |
| 2007 | Gaspar Bakos | Harvard-Smithsonian CfA | |
| 2007 | Rachel Kuzio de Naray | Univ. of California, Irvine | |
| 2007 | Joseph Gelfand | New York Univ. | |
| 2007 | Eric Hallman | Univ. of Colorado, Boulder | |
| 2007 | Joseph Hennawi | Univ. of California at Berkeley | |
| 2007 | John Johnson | Univ. of Hawaii | |
| 2007 | Marshall Perrin | Univ. of California, Los Angeles | |
| 2008 | Bethany Cobb | Univ. of California at Berkeley | |
| 2008 | David Lai | Univ. of California, Santa Cruz | |
| 2008 | Makenzie Lystrup | Univ. of Colorado, Boulder | |
| 2008 | Karín Menéndez-Delmestre | OCIW | |
| 2008 | Jeremiah Murphy | University of Washington | |
| 2008 | John Wisniewski | University of Washington | |